\documentclass[12pt]{iopart}

\usepackage[dvips]{graphicx}
\begin{document}

\title{The bounce hardness index of gravitational waves}

\author{Fumihiko Ishiyama$^1$ and Ryutaro Takahashi$^2$}

\address{$^1$
NTT Energy and Environment Systems Laboratories, 
Nippon Telegraph and Telephone Corp., Musashino, Tokyo 180-8585, JAPAN}
\address{$^2$National Astronomical Observatory of Japan, Mitaka, Tokyo 181-8588, JAPAN}

\ead{ishiyama.fumihiko@lab.ntt.co.jp}
\begin{abstract}
We present a method of mode analysis to search for signals with frequency evolution and limited duration
{
in a given data stream.
Our method is a natural expansion of Fourier analysis, and 
we can obtain the information about frequency evolution with high frequency precision and high time resolution.
}
Applications of this method to the analysis of in-spiral and burst signals show that 
the signals are characterized by an index which we name ``bounce hardness''.
The index corresponds to the growth rate of the signals.
\end{abstract}


\maketitle

\section{Introduction}
{
Interferometer gravitational wave detectors have been operated around the world 
to open a new window for astronomy\cite{Kawamura, Accadia, LIGO-LowMassBin79,LIGO1,LIGO-LowMassBin80,LIGO2,LIGO-GEO-Virgo}.
Though gravitational waves have still not been detected, 
upper limits for the wave signals such that from binary coalescence\cite{LIGO-LowMassBin79,LIGO1,LIGO-LowMassBin80} 
and from stellar core collapse\cite{LIGO2,LIGO-GEO-Virgo}
have been obtained by the operations.

To search for the signals, matched filtering,
which is a method with predicted waveforms,
is commonly used.
For example, 
exponential Gaussian correlator\cite{Clapson, Acernese} uses Gaussian envelope,
$\Omega$-pipeline\cite{omega-pipe} uses bi-square envelope,
and
coherent WaveBurst\cite{cWB} uses Mayer wavelets.
Some of them are used with a variation of linear predictive coding\cite{lpc},
which was originally developed for speech recognition.

Matched filtering 
is successfully used to analyze not only the chirp signals 
by binary coalescence but also the burst signals by core collapse\cite{Clapson},
where the waveforms caused by in-spiral binary stars are well anticipated by post-Newtonian calculations, 
and where there are various waveforms caused by stellar core collapse\cite{Ott} depending on the models of gravitational wave emission process.

In near future, the signals will be found by the matched filtering approach.
Once the signals are obtained with certain signal-to-noise ratio (SNR), 
a method for detailed analysis of the signals is required.
Obtaining the characteristics of the observed waves by the detailed analysis, 
more efficient signal detection comes to possible.

Therefore, we introduce our method of signal analysis for the detailed analysis.
}
Our method  is a kind of mode analysis, 
in which we fit a given time series 
with a linear combination of nonlinear oscillators. 
Our method with certain limitations is equivalent to Fourier analysis, 
and 
without the limitations, 
the frequency precision and the time resolution of the frequencies 
are beyond those of Fourier analysis.
For example, we can obtain the frequency information even from a fraction of cycle data, 
and we can obtain the characteristics of the data without oscillation,
{
as we show in the following sections.
}
As our method does not require 
{
specific assumptions
such like predicted waveforms,
}
it is applicable to the general field of signal analysis.

Through the analysis, we show  that
the waveforms of pre-bounce gravitational wave signals are characterized and categorized by an index $I$,
which we name ``bounce hardness''.
The index corresponds to the growth rate of the signals.

To begin with, we introduce our method of analysis and its simple implementation.
Then, we apply the method to the analysis of in-spiral and burst signals
{
without background noise,
}
and we show that the signals are characterized by an index of bounce hardness.
Finally, we summarize the paper.

\section{Method of analysis}

The method we introduce here is a kind of mode analysis.

{
Suppose that $g(t)$ is the output of a dynamical system which we want to analyze, 
and suppose that we can obtain $g(t)$ only as a numerical time series data.
Then, we fit the
}
time series $g(t)$ with 
a linear combination of oscillators with nonlinear oscillation.
The fitting function $h(t)$ is written as 
\begin{equation}
h(t) = \sum_{m=1}^M \exp H_m(t)
\end{equation}
where
$M$ is the number of oscillators
and
$H_m(t)$ is the complex function of the nonlinear oscillation
which is continuous and differentiable.

For a short enough time scale, the fitting function $h(t)$ is approximated by a linear function.
The linear function $\bar{h}_{0}(t) $ is written as
\begin{equation}
\bar{h}_{0}(t)  = \sum_{m=1}^M \exp \left[ H_m(t_0) + H^{\prime}_m(t_0) \left( t-t_0 \right) \right]
\end{equation}
for $t \sim t_0$. 
$H_m(t_0)$ and $H_m^\prime(t_0)$  correspond to the power/phase and the mode of the oscillation, respectively.

As the approximated function $\bar{h}_{0}(t)$ is a linear function, 
we can easily fit the given time series $g(t)$ with the function for $t \sim t_0$.

We show an  example of the fitting:
in general, the time series $g(t)$ is given as a discrete time series $g(t_0 + n \Delta T)$, 
where $n$ is the index of the time series and $\Delta T$ is the time interval of the sampling;
we fit the discrete time series with a linear difference equation 
\begin{equation}\label{eq-lde}
\prod_{m=1}^M \left\{  1 - \exp\left[ H_m^\prime (t_0) \Delta T \right] D \right\} g(t_0) = 0
\end{equation}
where  $D$ is the time shift operator such that
\begin{equation}
D^n g(t_0) = g(t_0 - n \Delta T) 
\end{equation}
and 
the mode of the oscillation 
$H_m^\prime (t_0)$ is obtained by solving the linear difference equation \eref{eq-lde}.

We introduce the following functions
\begin{eqnarray}
c_m(t_0) &=& H_m(t_0)
\\
f_m(t_0) &=& \frac{1}{\rmi 2 \pi} \Im  H^{\prime}_m(t_0) 
\\
\lambda_m(t_0) &=& \Re   H^{\prime}_m(t_0) 
\end{eqnarray}
to give physical meaning that 
the power/phase $c_m(t_0)$, the frequency $f_m(t_0)$, and the decay/growth rate $\lambda_m(t_0)$ at time $t=t_0$. 

By  use of the functions, the fitted function $h(t)$  for $t \sim t_0$ is written as 
\begin{eqnarray}
\left.  h(t) \right|_{t \sim t_0} &\simeq& \bar{h}_{0}(t)
\\
&=& \sum_{m=1}^M \exp \left\{  c_m(t_0) + \left[ \rmi 2 \pi f_m(t_0) + \lambda_m(t_0) \right] \left( t-t_0 \right) \right\}
\end{eqnarray}
and the time series of the oscillating mode is written as
\begin{equation}\label{eq-h-prime}
 H^{\prime}_m(t)  = \rmi 2 \pi f_m(t) + \lambda_m(t) .
\end{equation}

We note here the relation with discrete Fourier transform (DFT).
The following settings
\begin{eqnarray}
c_m(t) &=& C_m
\\
f_m(t) &=& \frac{m}{ M \Delta T} 
\\
\lambda_m(t) &=& 0
\\
M&=&N
\end{eqnarray}
where $N$ is the number of samples for the analysis,
gives DFT
\begin{equation}
h(t) = \sum_{n=1}^N \exp \left[  C_n + \rmi 2 \pi \frac{n}{N \Delta T}  \left( t-t_0 \right) \right]
\end{equation}
itself.
Therefore, it is found that our method of analysis includes traditional Fourier analysis.

The major characteristic of our method of analysis is that 
we can obtain the time series of the oscillating modes \eref{eq-h-prime} of given time series with high time resolution.
Each mode contains the information about frequency and decay/growth rate.
In contrast, Fourier analysis is a method which has fixed modes and fitting parameter $C_n$. 
Therefore, time/frequency resolution is limited by the number of samples for analysis, 
and decay/growth rates are not obtained directly.

\section{Implementation}

In this section, 
we introduce a simple implementation of our method of analysis
to  apply the method to the analysis of gravitational wave signals.
The precision and time resolution of our method of analysis are shown through the application.

We employ least-mean-square (LMS) fitting for the analysis in this paper.
The simultaneous equations for the fitting are 
\begin{equation}\label{eq-lms-fit}
\frac{\partial}{\partial a_{m^\prime}} \sum_{n=1-N}^{0} \left[ \left( 1-\sum_{m=1}^M a_m D^m \right) g(t_0+n \Delta T) \right]^2   = 0
\end{equation}
for $m^\prime=1, 2, ..., M$, 
where $a_m$ is the coefficient of the linear difference equation to be fitted, 
and 
$N$ is the number of samples for the LMS fitting.
We obtain $a_m$ by solving the simultaneous equations.
Then, the oscillating mode $H_m^\prime(t_0)$ is obtained by the factorization
\begin{equation}
1-\sum_{m=1}^M a_m D^m = \prod_{m=1}^M \left\{  1 - \exp\left[ H_m^\prime (t_0) \Delta T \right] D \right\} .
\end{equation}
We solve the simultaneous equations for 
\begin{equation}
t_n = t_0 + n \Delta T ,
\end{equation}
and we obtain the time series of the oscillating mode $H_m^\prime(t_n)$.

{
We note here that this implementation corresponds to the traditional maximum entropy method (MEM)
when $M$ and $N$ are large enough.
Therefore, background random noise, whose time series can not be written by a linear equation, 
is weeded out as the residual of the LMS fitting.
In addition, 
we notice here that we should solve \eref{eq-lms-fit} exactly.
It is for the reason that 
the traditional method of the LMS fitting contains implicit periodic boundary condition
to reduce computational complexity, 
and it gives wrong solution.
}

\section{Application to chirp signal}

In the following, we apply the above implementation to the analysis of chirp signal.

\begin{figure}[hbtp]
\begin{center}
\includegraphics{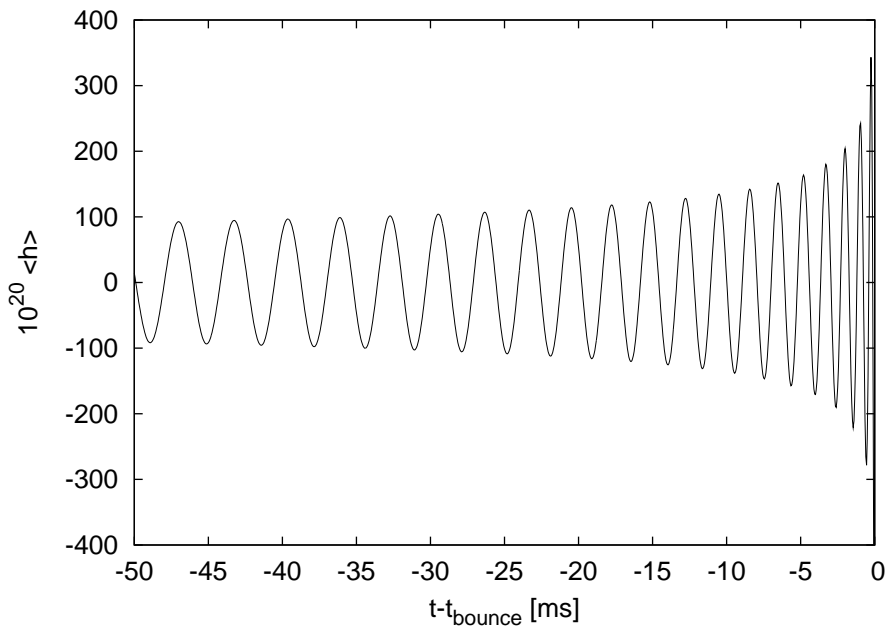}
\end{center}
\caption{
Time series of chirp signal \eref{eq-g-chirp} for the analysis.
}
\label{fig-ch-ts}
\end{figure}

We employ the chirp signal of  in-spiral binary stars with mass $1.4 M_\odot$ 
and 
the distance 10 kpc for the analysis \cite{Schutz}.
The equations for the chirp frequency $f_c(t)$ and the chirp signal $h_c(t)$ are
\begin{eqnarray}
f_c(t) &=& \frac{1}{\pi}  \left( \frac{c^3}{G} \right)^{5/8}  \left( \frac{5}{256 \mu M_{\rm tot}^{2/3} } \frac{1}{t_{\rm bounce} -t} \right) ^{3/8}
\\
h_c(t) &=& 4 \frac{ G^{5/3} } {c^{4} }  \frac{ \mu} {r} \left( \pi M_{\rm tot} f_c(t) \right) ^{2/3} \sin \left[ 2 \pi f_c(t) \left( t - t_{\rm bounce} \right) \right] 
\label{eq-h}
\end{eqnarray}
where
$M_{\rm tot}=2.8M_\odot$,
$\mu=0.7 M_\odot$,
and
$r=10$ kpc.
The time series for the analysis $g(t)$ becomes
\begin{equation}\label{eq-g-chirp}
g(t_n) = h_c(n \Delta T + t_{\rm bounce})
\end{equation}
where $\Delta T$ is the time interval of the sampling,
and we choose 0.05 ms as  $\Delta T$.
We show the time series \eref{eq-g-chirp}  in \fref{fig-ch-ts}.

As the chirp signal $h_c(t)$ contains a single function for frequency, 
we choose 2 as the number of nonlinear oscillators $M$.
Then, the chirp signal $h_c(t)$ is written as 
\begin{equation}
h_c(t) = \exp H_+(t) + \exp H_-(t) 
\end{equation}
where the complex functions $H_\pm(t)$ of the nonlinear oscillators are written as
\begin{equation}
H_\pm(t) = -\frac{1}{4} \ln \left( t_{\rm bounce} -t \right) \pm \rmi 2 \pi \cdot 133.5 \left( t_{\rm bounce} -t \right) ^{5/8} -43 \mp \frac{\pi}{2} \rmi.
\end{equation}

In these conditions,
the frequency time series $f(t)$ of the nonlinear oscillators  becomes
\begin{eqnarray}
 f(t) &=&
  \frac{1}{2 \cdot \rmi 2 \pi } \left(H_+^{\prime}(t) - H_-^{\prime}(t) \right) 
\\
&=& 133.5 \cdot \frac{5}{8} \left( t_{\rm bounce} - t \right)^{-3/8} 
\label{eq-f}
\end{eqnarray}
and the growth rate time series $\lambda(t)$ of the nonlinear oscillators   becomes 
\begin{eqnarray}\label{eq-l-correction}
 \lambda(t) 
&=&
 \frac{1}{2} \left(H_+^{\prime}(t) + H_-^{\prime}(t) \right) 
+ \frac{1}{2} 
\frac{
 H_+^{\prime\prime}(t) - H_-^{\prime\prime}(t) 
}{
 H_+^{\prime}(t) - H_-^{\prime}(t) 
}
\\
&=& -\frac{7}{16} \left( t_{\rm bounce} - t \right)^{-1} .
\label{eq-l}
\end{eqnarray}
Here, the second term in  \eref{eq-l-correction} is a correction term.
The term is required because the complex functions $H_\pm(t)$  have strong nonlinearity,
and locally linearized functions are different from the complex functions themselves.

\begin{figure}[hbtp]
\begin{center}
\includegraphics[scale=1]{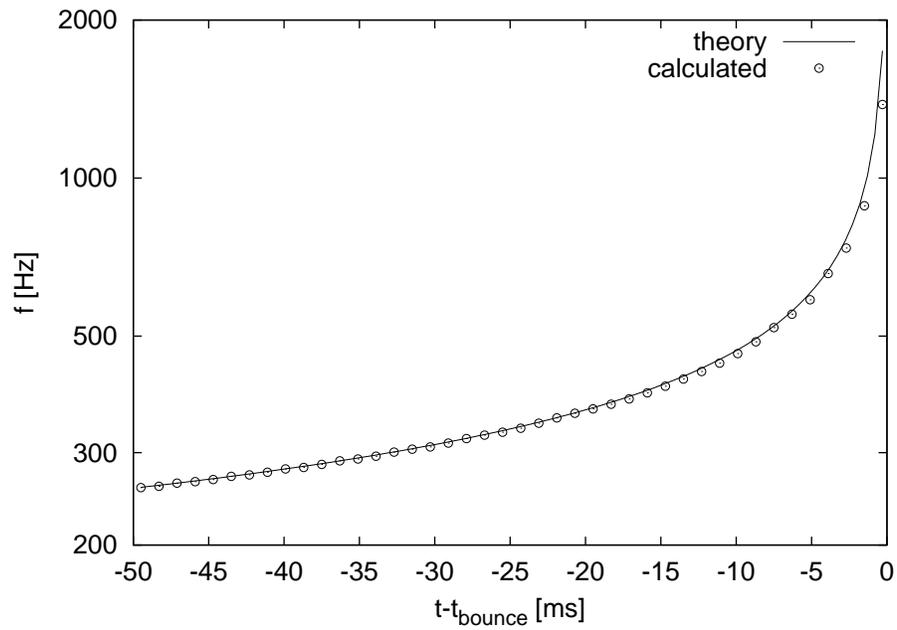}
\end{center}
\caption{
Semi-log plot of chirp frequency.
Theoretically expected frequency \eref{eq-f} is shown by solid line, 
and calculated one from time series \eref{eq-g-chirp} is plotted by circles.
}
\label{fig-ch-f}
\end{figure}

\begin{figure}[hbtp]
\begin{center}
\includegraphics[scale=1]{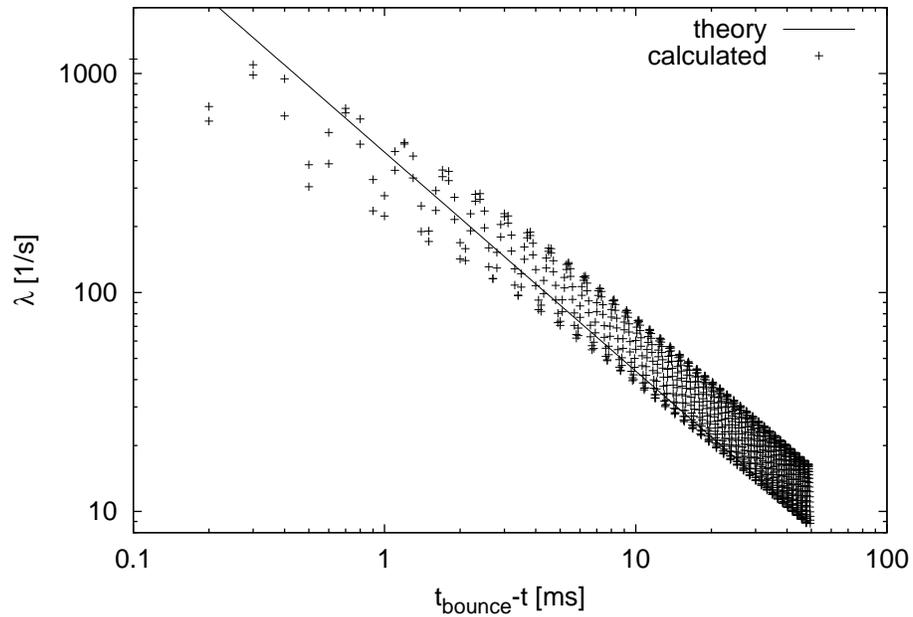}
\end{center}
\caption{
Log-log plot of chirp growth rate.
Theoretically expected growth rate  \eref{eq-l} is shown by solid line, 
and calculated one from time series \eref{eq-g-chirp} is plotted by plus signs.
}
\label{fig-ch-l}
\end{figure}

We choose 8 as the number for the LMS fitting $N$,
and we plot the results of the analysis 
in  \fref{fig-ch-f} (chirp frequency $f(t)$) 
and 
in \fref{fig-ch-l} (growth rate $\lambda(t)$).
As we employed the parameters $M=2$ and $N=8$, 
the number of  samples for calculating $f(t)$ and $\lambda(t)$ 
becomes 10 ($M+N$).
In addition, as we employed 0.05 ms as  $\Delta T$,
the time width of a single analysis becomes 0.5 ms.

\Fref{fig-ch-f}  shows the time series of frequency with a semi-log scale.
The solid line shows the theoretical time series of frequency \eref{eq-f}, 
and the plots by circles show the calculated one from time series \eref{eq-g-chirp}.
Circles are plotted every 12 ms to maintain readability.
The calculated time series of frequency agrees strongly with the theoretical one.
{
The difference between them is less than $2\%$ for $t-t_{\rm bounce} < -10$, 
and is less than $10 \%$ for $t-t_{\rm bounce} < -1$.
As we assume a linear equation for a duration of 0.5 ms, 
the difference becomes larger approaching to the bounce.
}

\Fref{fig-ch-l}  shows the time series of growth rate with a log-log scale.
The solid line shows the theoretical time series of growth rate  \eref{eq-l}, 
and the plots by plus signs show the calculated one from time series \eref{eq-g-chirp}.
The trend and the values of the calculated time series of the growth rate agree fairly well with the theoretical one.
The oscillations observed in the time series of calculated values come from the LMS fitting.
As our time width of analysis (0.5 ms) is much shorter 
than the cycle of oscillation of the chirp signal ($\sim$5 ms) in \fref{fig-ch-ts}, 
the fitted value is affected by the status of the curvature of the chirp signal.
{
We can attenuate the amplitude of the oscillations by employing large enough  $N$ 
compare to the period of the oscillations of the time series.
However, as we assume a linear equation for a duration of $\left( M+N \right) \Delta T$,
it results low time resolution for tracing non-linearity.
}

As shown above, 
our method of analysis gives information about frequency and growth rate with high precision and with high time resolution.

\section{Application to burst signals}

In this section, we employ the signals of gravitational waves calculated and presented by Dimmelmeier et al. \cite{dimA,dimB}
for the analysis.
{
Among the signals, we employ the ones named A1B3G3 and A3B5G4.
They represents two kinds of signal growth, 
and 
the other signals are categorized in either group.
}

We prepared their time series data \cite{dimData} with 8 kHz sampling, 
and we employ the same implementation and the same parameters ($M=2$ and $N=8$) used in the previous section,
because it is the most simple implementation.
Then, we set the time series for the analysis $g(t)$ as 
\begin{equation}
g(t) = A_{\rm 20}^{\rm E2} 
\end{equation}
which is given by Dimmelmeier et al.,
and we fit the time series with 
\begin{equation}
h(t) = \exp H_1(t) + \exp H_2(t) .
\end{equation}

As the pre-bounce time series has no oscillation, 
the imaginary part of $H_m(t)$ becomes zero,
and we obtain two growth rates $\lambda_1(t)$ and $\lambda_2(t)$.

Through the analysis, 
we found that their time series are characterized by an index of bounce hardness $I$, 
and the time series of growth rate  $H^\prime_m(t)$ is written as
\begin{equation}\label{eq-index}
H_m^\prime(t) = \lambda_m(t) \propto \left(t_{\rm bounce} -t \right)^{-I} .
\end{equation}
Therefore, the fitted time series $h(t)$ of gravitational wave becomes
\begin{equation}\label{eq-h-cdI}
h(t) = \sum_{m=1}^2 \exp \left[ c_m + d_m \left(t_{\rm bounce} -t \right)^{1-I} \right] 
\end{equation}
where $c_m$ and $d_m$ are the fitting parameters for each time series, 
except when $I=1$.
The case $I=1$ corresponds to the case in the previous section \eref{eq-l},
where the real part of $H_m(t)$ is logarithmic.

{
The signals with  $I \leq 0$ has no singularity at the bounce.
In contrast, 
$I > 0$ has  singularity in the gradient, and $I \geq 1$ has singularity in the amplitude.
In this way, the index $I$ represents the hardness of the bounce.
}

In the following, we show the results of the analyses for A1B3G3 and A3B5G4,
the harder case and  softer case, respectively.
{
In addition, we categorize the other signals in the catalog \cite{dimData}
by using the index $I$.
}

\subsection{A1B3G3}

\begin{figure}[hbtp]
\begin{center}
\includegraphics[scale=1]{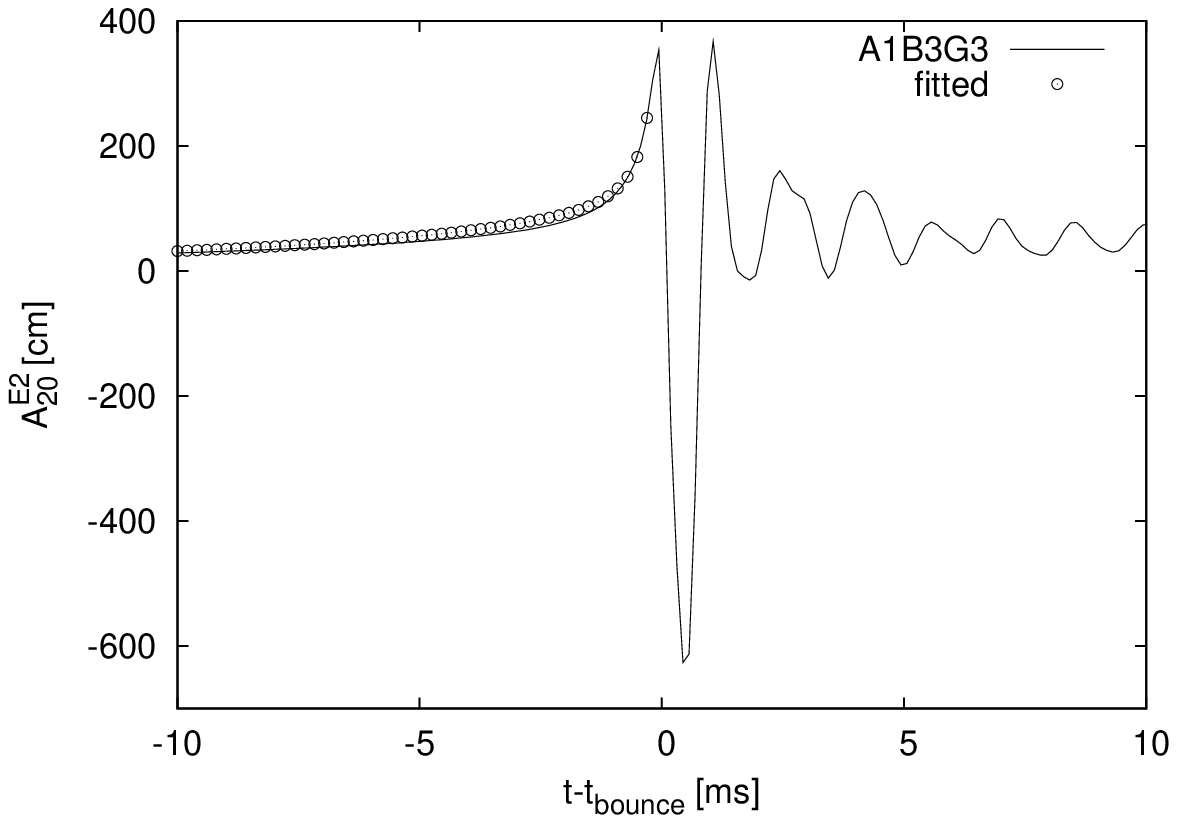}
\end{center}
\caption{
Time series of A1B3G3 is shown by solid line, 
and our fitted time series \eref{eq-133-h} is plotted by circles.
}
\label{fig-133-ts}
\end{figure}

\begin{figure}[hbtp]
\begin{center}
\includegraphics[scale=1]{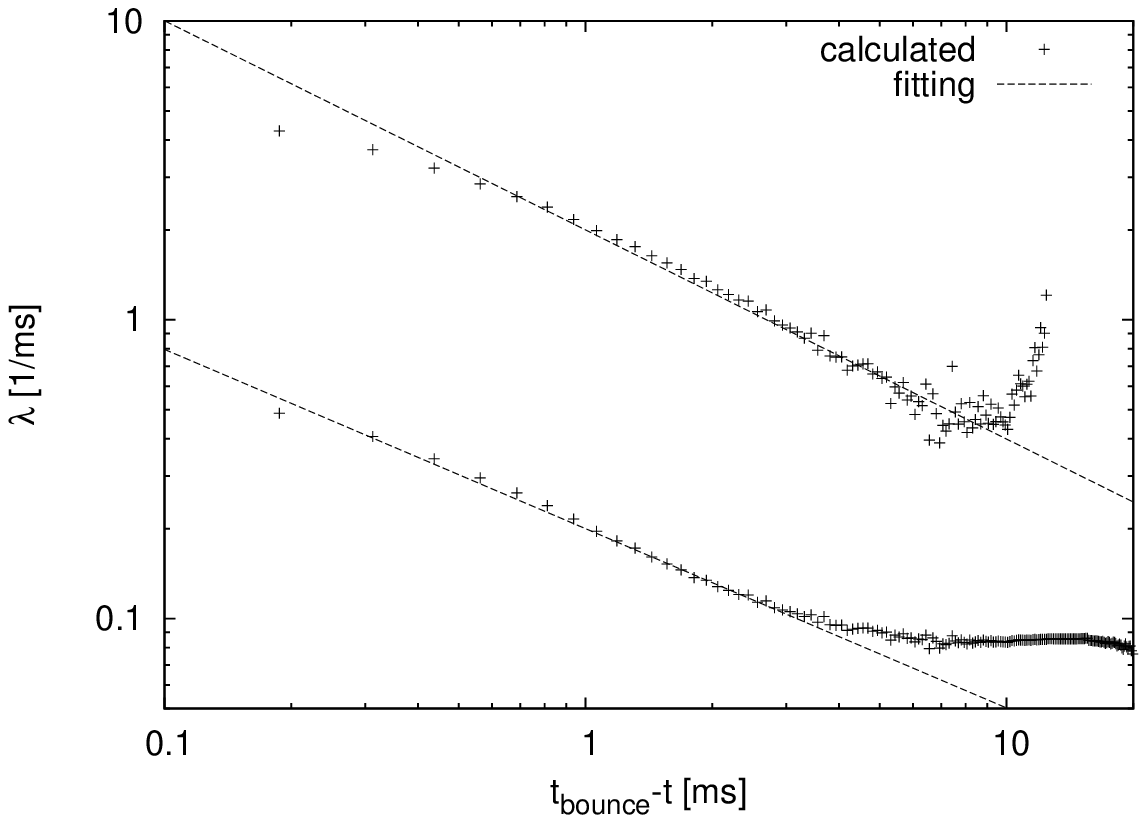}
\end{center}
\caption{
Log-log plot of growth rate $\lambda$. 
Two lines shown by plus signs imply polynomial functions.
Two fitted functions \eref{eq-133-l1} and \eref{eq-133-l2} are shown by dashed lines.
}
\label{fig-133-l}
\end{figure}

\Fref{fig-133-ts} shows the time series of A1B3G3 and the fitted time series.
\Fref{fig-133-l} shows the plots of  growth rate $\lambda_m(t)$ in a log-log scale.

The plots of growth rates in \fref{fig-133-l} show two lines, 
and we fit them with straight lines.
By the fitting, we obtain the following two functions of growth rates
\begin{eqnarray}
\lambda_1(t) &=& 2.3 (t_{\rm bounce}-t)^{-0.4} \label{eq-133-l1}
\\
\lambda_2(t) &=& 0.25 (t_{\rm bounce}-t)^{-0.4} \label{eq-133-l2}
\end{eqnarray}
and the fitted function in \fref{fig-133-ts} 
\begin{equation} \label{eq-133-h}
h(t)=702 \exp[-\frac{2.3}{0.6} (t_{\rm bounce}-t)^{0.6} ]
	+ 168 \exp[-\frac{0.25}{0.6} (t_{\rm bounce}-t)^{0.6} ] 
\end{equation}
where we employed the obtained values 
at $t=t_{\rm bounce}-0.4375$ to set the parameters $c_m$ and $d_m$ in \eref{eq-h-cdI}.
{
The difference between A1B3G3 and \eref{eq-133-h} is less than $10 \%$ 
for  $-1.4 < t-t_{\rm bounce} < -0.1$, 
and the maximum value of the difference for  $-10 < t-t_{\rm bounce} < -1.4$ is $20 \%$ at  $t-t_{\rm bounce} = -3.5$.
}

In this case, the index of bounce hardness becomes $I=0.4$,
and the gradient of the time series has singularity at the bounce.

\subsection{A3B5G4}

\Fref{fig-354-ts} shows the time series of A3B5G4 and the fitted time series.
\Fref{fig-354-l} shows the plots of  growth rate $\lambda_m(t)$ in a log-log scale.

In contrast to the A1B3G3 case, the two lines plotted by plus signs in \fref{fig-354-l} 
are independent of time.
This implies that $\lambda_m(t)$ is a constant function.
We fit them with straight lines, and we obtain growth rates
\begin{eqnarray}
\lambda_1 &=& 0.824 \label{eq-354-l1}
\\ 
\lambda_2 &=& 0.108 \label{eq-354-l2}
\end{eqnarray}
and the fitted function in \fref{fig-354-ts} 
\begin{equation} \label{eq-354-h}
h(t)=620 \exp [0.108 (t-t_{\rm bounce}) ]
	-140 \exp [0.824(t-t_{\rm bounce})  ]
\end{equation}
where we employed the obtained values 
at $t=t_{\rm bounce}-0.0625$ to set the parameters $c_m$ and $d_m$ in \eref{eq-h-cdI}.
{
The difference between A3B5G4 and \eref{eq-354-h} 
is less than $2 \%$ for  $-10 < t-t_{\rm bounce} < 0$,
and
is less than $4 \%$ for  $-22.5 < t-t_{\rm bounce} < 1$.
}

In this case, the index of bounce hardness becomes $I=0$,
and the gradient of the time series has no singular points
during the bounce.

{

\begin{figure}[hbtp]
\begin{center}
\includegraphics[scale=1]{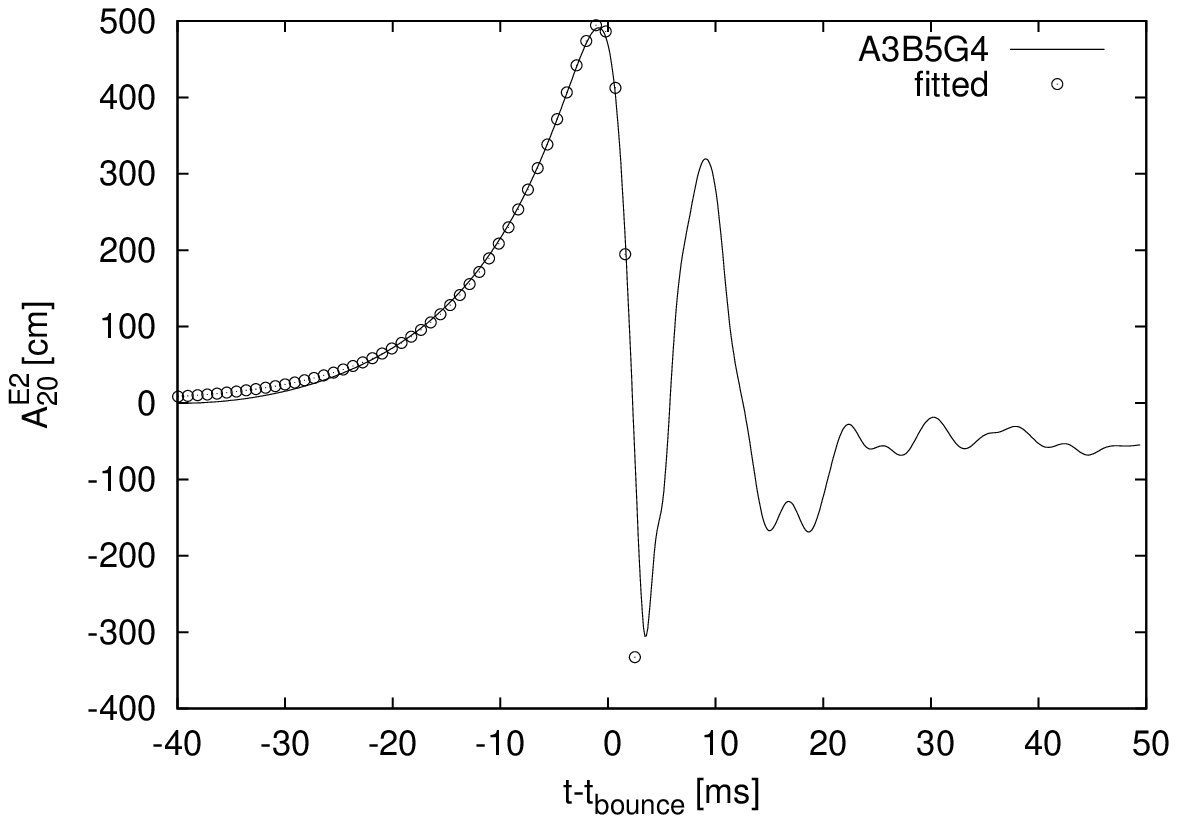}
\end{center}
\caption{
Time series of A3B5G4 is shown by solid line, 
and our fitted time series \eref{eq-354-h} is plotted by circles.
}
\label{fig-354-ts}
\end{figure}

\begin{figure}[hbtp]
\begin{center}
\includegraphics[scale=1]{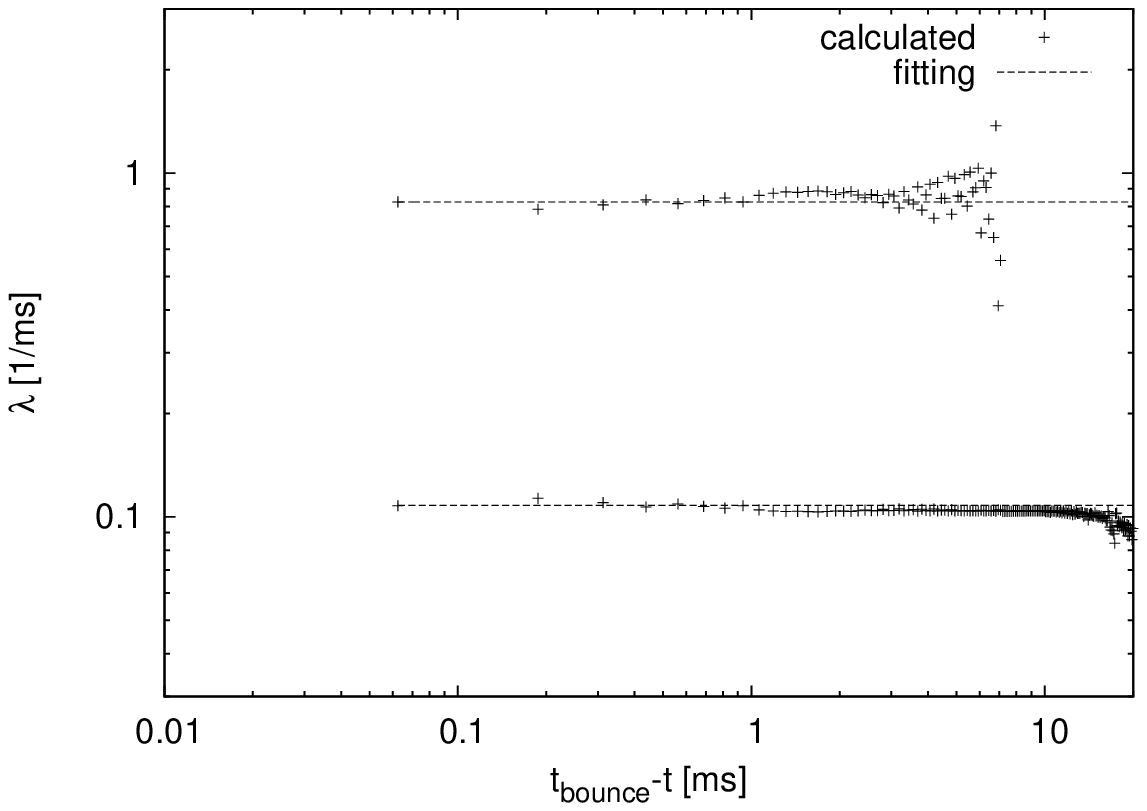}
\end{center}
\caption{
Log-log plot of growth rate $\lambda$.
Two lines shown by plus signs imply constant functions.
Two fitted functions \eref{eq-354-l1} and \eref{eq-354-l2} are shown by dashed lines.
}
\label{fig-354-l}
\end{figure}

\subsection{Categorizing signals}

We show the index  $I$
of each signal in the catalog\cite{dimData},
and that of in-spiral binary stars in Table \ref{tab-I}.

\begin{table}[hbt]
\caption{\label{tab-I} Bounce hardness index  $I$  of  signals.}
\begin{tabular*}{\textwidth}{@{}l*{15}{@{\extracolsep{0pt plus
12pt}}l}}
\br
$I$ & Signal growth  & Examples of  signals \\
\mr
$I=0 $ & Exponential & A2B4G1, A3B4G2, A3B5G4, A4B4G4, A4B4G5, A4B5G5 \\

$0<I<1$ & Intermediate & A1B3G3, A1B3G5, and most of the other signals \\

$I =1$ & Polynomial & In-spiral binary stars \\
\br
\end{tabular*}
\end{table}

The table shows that the rapidly (large $\beta_{\rm rot \ ini}$) and 
highly differentially (small $A$) rotating initial models exhibit exponential signal growth.
Among them, the growth rate  of A2B4G1 shows some fluctuation around a value, 
and it is not a simple constant.
Therefore, it seems that a small enough $A$ is required for stable exponential signal growth.

Most of the other burst signals exhibit intermediate (between exponential and polynomial) signal growth,
and the few resting signals are not identified simply.

}

\section{Conclusions}

{
We introduced our own method of signal analysis,
which is a natural expansion of Fourier analysis.
Using our method, high resolution information about frequency and decay/growth rate
is obtained from small number of samples.
Therefore, the detailed analysis of frequency evolution and decay/growth rate evolution 
comes to possible.
}

We applied the method to the analysis of the chirp signal
and the burst signals of gravitational waves.
We found that the signals are characterized by an index
of bounce hardness $I$.
{
The growth rate $\lambda_m(t)$ of the signals is written as
\eref{eq-index}
and the index  $I$ is 
$1$ for the chirp signal (polynomial signal growth),
$0.4$ for A1B3G3 (intermediate signal growth), 
and
$0$ for A3B5G4 (exponential signal growth).
The rapidly and highly differentially rotating initial models belong to A3B5G4 group, 
and 
most of the other burst signals belong to  A1B3G3 group.
}

We note here that this is the first paper introducing our method of analysis, 
and there are no references on the method.
In addition, the method of analysis is still under development, 
and more qualified analysis will become possible through the progress of the development.
{
For example, 
the way to detect signals from low SNR data stream by using our method of analysis, 
and
the way to apply our method of analysis into matched filtering
will be found.

}

\end{document}